# Metallic state in the vicinity of molecular orbital crystallization in the $d^1$ thiospinel ZnTi$_2$S$_4$ prepared via a reductive ion-exchange reaction

Yuya Haraguchi[1,*], Hiroki Arikai[1], and Hiroko Aruga Katori[1,2]

[1]Department of Applied Physics and Chemical Engineering, Tokyo University of Agriculture and Technology, Koganei, Tokyo 184-8588, Japan
[2]Research Center for Thermal and Entropic Science, Graduate School of Science, Osaka University, Toyonaka, Osaka 560-0043, Japan

E-mail: chiyuya3@go.tuat.ac.jp

**Abstract**

A novel Ti$^{3+}$-based thiospinel ZnTi$_2$S$_4$ is successfully synthesized via a low-temperature ion-exchange reaction. ZnTi$_2$S$_4$ shows a signature of metallic ground state evidenced by a contribution of conduction electrons in the heat capacity and Pauli-like paramagnetic susceptibility. These observations contrast to the electronic state of similar Ti$^{3+}$-based spinel MgTi$_2$O$_4$ exhibiting the metal-insulator transition associated with a molecular orbital crystallization (MOC). Furthermore, the magnetic susceptibility of ZnTi$_2$S$_4$ shows a pseudogap-like behavior indicated by a vast peak in the magnetic susceptibility around 110 K, likely originating from the MOC fluctuation. The origin of the difference in the electronic states of MgTi$_2$O$_4$ and ZnTi$_2$S$_4$ would be due to the different magnitude of overlap between Ti $3d$ and $p$ orbitals (O: $2p$ and S: $3p$). The presence of a MOC state in the close vicinity of insulator-metal transition may suggest the importance of itinerancy in a MOC.

## I. Introduction

In condensed matter physics, the importance of self-organization phenomena due to the entanglement of electrons with multiple degrees of freedom has been discussed [1-15]. For example, the spontaneous molecular cluster formation is driven by molecular orbitals with strong metal-metal bonding between adjacent transition metal ions and the periodic modulation in a crystal. Such a phenomenon, in which multiple degrees of freedom of electrons form metal-metal bond ion clusters and arrange them throughout the lattice, is termed molecular orbital crystallization (MOC). The importance of MOCs instability in dramatic phenomena such as metal-insulator transitions with crystal transformation has been discussed recently [1].

The diversity of cluster patterns in MOCs is strongly coupled to lattice and electron degrees of freedom. For example, in a group of compounds with a simple transition metal ion network, such as the rutile-type structure, similar dimerization has been observed regardless of the type of ion. On the other hand, in frustrated lattices such as triangular and pyrochlore lattices, various types of the cluster are formed. In the two-dimensional triangular lattice system, trigonal trimers are formed in LiVO$_2$ [2,3] and LiVS$_2$ [4,5]. In contrast, a ribbon chain of one-dimensional clusters consisting of multiple linear trimers is formed in $M$Te$_2$ ($M$ = V, Nb, Ta) [6,7] and $Li_{0.33}VS_2$ [8]. As a more complex case, star-of-David-shaped clusters containing 13 atoms are formed in $TaS_2$ [9,10]. In pyrochlore lattice, helically arranged Ti-Ti dimers are formed in $MgTi_2O_4$ [11,12], heptamers in $AlV_2O_4$ [13,14], and octamers in $CuIr_2S_4$ [15]. Thus, it is expected that the search for new materials in which ions with MOC instability are placed on the frustrated lattice will lead to the discovery of new electronic states.

The MOC instability sometimes competes with the itinerancy of electrons. For example, $LiVSe_2$ is metal while $LiVS_2$ exhibits MOC transition due to reduced itinerancy [4]. Furthermore, the solid solution system $LiVS_{2-x}Se_x$ shows a signature of a pseudogap formation as seen in copper oxides, which has been interpreted as a precursor phenomenon to MOC [4]. The replacement of anions by other larger anions (for example, O → S → Se) is considered to enhance the overlap between $3d$ and $p$ electrons, which results in broadening the bandwidth. Thus, these changes in physical properties in the $LiVS_{2-x}Se_x$ system are attributed to the enhancement of itinerancy. Moreover, the MOC instability is strongly related to the number of $d$ electrons. For example, no MOC transition appears in a three-dimensional spinel system with $d^{0.5}$-electrons, such as the spinel $LiTi_2O_4$ and thiospinel $CuTi_2S_4$, contrasting with the MOC ground state in $MgTi_2O_4$ with $d^1$-electrons. Thus, the hypothetical $d^1$ thiospinel is expected to exhibit novel electronic properties due to the intense competition between the MOC instability and the itinerancy of electrons. However, no synthesis of $d^1$ thiospinel has been reported so far.

Here, we report the successful synthesis of a new $Ti^{3+}$ thiospinel $ZnTi_2S_4$ using a topochemical reaction of $CuTi_2S_4$ with Zn metal and its physical properties. The electronic state of $ZnTi_2S_4$ is metallic, evidenced by a contribution of conduction electrons in the heat capacity at low temperatures and poor temperature dependence of the electric resistivity and magnetic susceptibility, which contrasts to similar $d^1$ spinel $MgTi_2O_4$ with a MOC ground state.

## II. Experimental Procedure

According to the previous report, the precursor $CuTi_2S_4$ was obtained by the conventional solid-state reaction method [16]. This precursor was ground well with fourfold excess of Zn metal, sealed in an evacuated Pyrex tube, and reacted at 380°C for 100 h as follows,

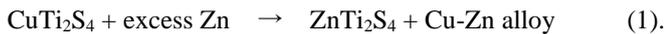

$CuTi_2S_4$ + excess Zn → $ZnTi_2S_4$ + Cu-Zn alloy     (1).

The byproduct of Cu-Zn alloy was removed by washing the sample with aqueous ammonia. The product was characterized by powder X-ray diffraction (XRD) experiments in a diffractometer with Cu-Kα radiation. The cell parameters and crystal structure were refined by the Rietveld method using the RIETAN-FP v2.16 software [17].

The temperature dependence of the magnetic susceptibility was measured using a magnetic property measurement system (MPMS; Quantum Design) equipped at the Institute for Solid State Physics at the University of Tokyo. The temperature dependence of the electric resistivity was measured using a conventional four-probe method with a hand-made apparatus. The temperature dependence of the specific heat was measured using a relaxation method with a physical property measurement system (PPMS; Quantum Design) equipped at the Institute for Solid State Physics at the University of Tokyo.

**Table 1** Crystallographic parameters for $ZnTi_2S_4$ (space group: $Fd$-$3m$) determined from powder x-ray diffraction experiments. The obtained cubic lattice parameters are $a = 10.1355(1)$ Å. $B$ is the atomic displacement parameter and $g$ is the occupancy.

| atom | g | x | y | z | B(Å) |
|---|---|---|---|---|---|
| Zn | 1 | 1/8 | 1/8 | 1/8 | 1.27(2) |
| Ti | 1 | 1/2 | 1/2 | 1/2 | 0.49(2) |
| S | 1 | 0.25696(5) | = x | = x | 0.62(1) |

## III. Results

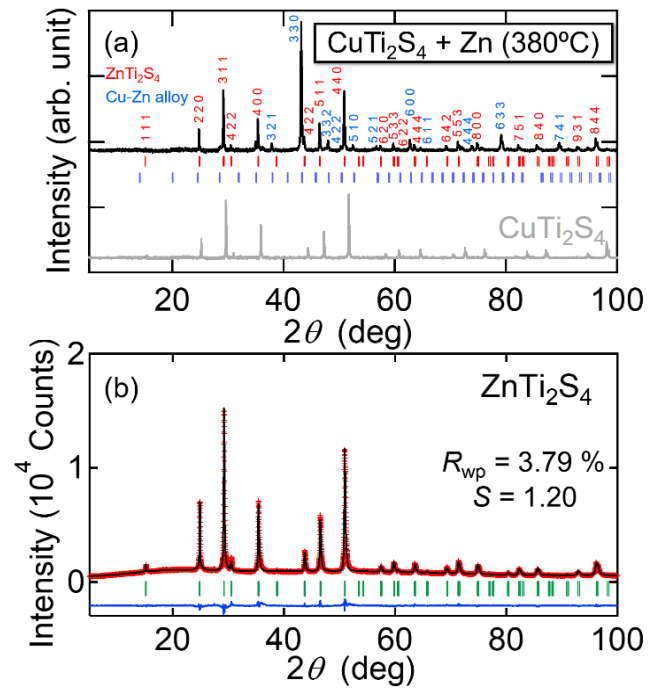

**Figure 1** (a) Powder x-ray diffraction patterns of precursor $CuTi_2S_4$ and the product from $CuTi_2S_4$ + Zn reaction. Red and blue vertical bars with peak indices are given for cubic $Fd$-$3m$ ($ZnTi_2S_4$) and $I$-$43m$ (byproduct Cu-Zn alloy) unit cells with the lattice constants $a = 10.136$ and $8.864$ Å, respectively. (b) Powder x-ray diffraction patterns of $ZnTi_2S_4$ after washing with aqueous ammonia. The observed intensities (red), calculated intensities (black), and their differences (blue) are shown. Vertical bars indicate the positions of Bragg reflections.

A powder XRD pattern of as-synthesized product is shown in Fig. 1(a). For comparison, the powder XRD pattern of precursor $CuTi_2S_4$ is also shown. It is observed that the lattice parameter increases (9.999 Å → 10.136 Å) with maintaining the spinel structure after the reaction. Besides, extra peaks derived from the byproduct Cu–Zn alloy (γ-brass) with a space group of *I-43m* are observed, indicating that the reaction (1) has proceeded. When the reaction was carried out at higher temperatures than 400°C, ZnS was precipitated as impurities. This fact indicates that $ZnTi_2S_4$ is a metastable phase. Aqueous ammonia can easily remove the byproduct Cu-Zn alloy to get a pure sample of $ZnTi_2S_4$. The powder XRD pattern is perfectly reproduced using the spinel structure with the obtained cubic lattice parameters $a$ = 10.1355(1) Å as shown in Fig. 1(b). The details of the refinement parameters are given in Table 1. The chemical composition was examined using energy dispersive x-ray spectrometry (EDX), and it was found that Zn: Ti: S = 1.00(2): 2.00(2): 4.00(2), which is in good agreement with the ideal composition of $ZnTi_2S_4$. The average Ti-S bond lengths of $TiS_6$ octahedra are 2.4028 Å and 2.4654 Å for $CuTi_2S_4$ and $ZnTi_2S_4$, respectively, suggesting that the Ti-valence is effectively reduced by ion exchange. The bond valence sum calculation for Ti ions from the refined structural parameters [18] yields +3.525 and +2.977 for $CuTi_2S_4$ and $ZnTi_2S_4$, respectively, which are in excellent agreement with the expected valences.

Figure 2 shows the temperature dependence of electric resistivity ρ of a cold-pressed pellet $ZnTi_2S_4$. The room temperature resistivity is ~1 Ωcm, which is typical of a semiconducting state. Furthermore, in all temperature ranges, ρ increases substantially on cooling without an obvious transition. It shows, however, only a factor of 1.3 increase between 300 and 5 K, which would be too weak to be a conventional semiconductor. Moreover, as shown in the right inset of Fig. 2, ρ does not follow an activation behavior for an extensive temperature range. A forcedly Arrhenius fitting in the 200–300 K range yields a gap of $\Delta$ = 3.01(1) meV ~ 35 K. The estimated tiny gap must not generate a semiconducting resistivity ($d\rho/dT$ < 0) near room temperature. Therefore, it is reasonable to assume that this semiconducting behavior is not intrinsic. In order to interpret these data quantitatively, a grain-boundary effect must be considered [19]. Since this sample is a cold-pressed pellet, one would expect a uniform increase in resistivity due to a grain boundary scattering. We note that the semiconducting properties have been observed in sintered/cold-pressed polycrystalline samples, whereas metallic properties have been found in single crystals, for example, $CaCrO_3$ [20, 21] and $CrSe_2$ [22, 23]. As discussed in the following paragraphs about the heat capacity and magnetic susceptibility, a metallic ground state of $ZnTi_2S_4$ is demonstrated. Thus, the intrinsic resistivity of $ZnTi_2S_4$ may be much smaller and even metallic.

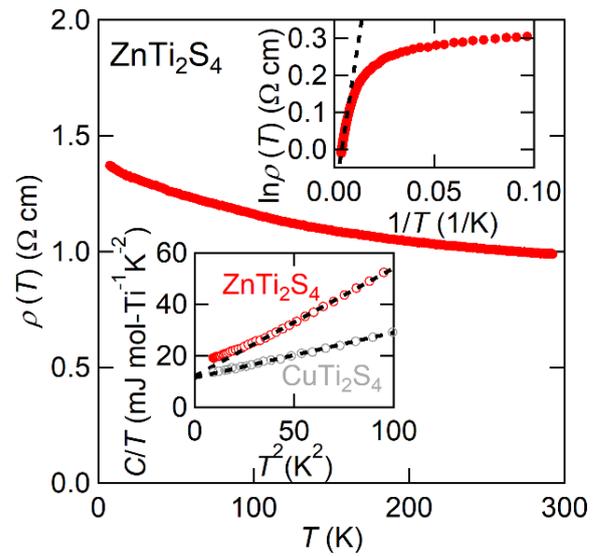

**Figure 2** The temperature dependence of electric resistivity ρ of a powder compact of $ZnTi_2S_4$. The right inset shows the Arrhenius plot, where the dashed line on the data represents a fit to thermally activated Arrhenius relationship yielding a gap of $\Delta$ = 3.01(1) meV. The left inset shows a specific heat divided by temperature $C/T$ as a function of $T^2$ below 10 K with a fit to the equation $C/T = \gamma + \beta T^2$ (dotted line). The $C/T$ data for $CuTi_2S_4$ are also shown for comparison.

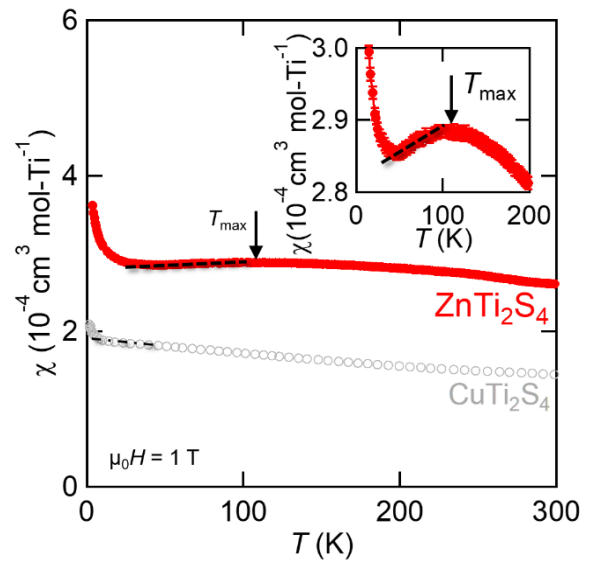

**Figure 3** The temperature dependence of magnetic susceptibility χ of a powder sample of $ZnTi_2S_4$. For comparison, χ of precursor $CuTi_2S_4$ is also shown. The inset shows the enlarged view of a vast peak in χ around $T_{max}$ ~ 110 K. The dashed line indicates a guide to the eye of a χ-decreasing below $T_{max}$.

To confirm the electrons are itinerant or not, we conducted the temperature dependence of heat capacity of $ZnTi_2S_4$. The low-temperature heat capacity divided by temperature $C/T$ exhibits a linear dependence with a finite ordinate intercept against $T^2$ in the temperature range between 2 and 10 K, as shown in the left inset of Fig. 2. The $C/T$ data were fit to the equation $C/T = \gamma + \beta T^2$, where $\gamma$ and $\beta$ model the electronic and phonon contributions to $C$, respectively. The $\gamma$ value of 12.116 mJ mol-Ti$^{-1}$K$^{-2}$ indicates a finite density of states (DOS) at the Fermi level. This observation strongly evidences that $ZnTi_2S_4$ is metallic. For comparison, the $C/T$ data of $CuTi_2S_4$ is also shown. The $\gamma$ value of 11.95 mJ mol-Ti$^{-1}$K$^{-2}$ is almost identical to $ZnTi_2S_4$, indicating that DOS is almost unchanged by electron doping. The Debye temperature $\Theta_D$ is estimated to be 253 K from $\beta$ = 0.420(3) mJ mol-Ti$^{-1}$K$^{-4}$, consistent with a typical value in materials with metallic conductivity.

Figure 3 shows magnetic susceptibility $\chi$ of the powder sample of $ZnTi_2S_4$ measured at magnetic fields of 1 T. The almost temperature-independent $\chi$ suggests a Pauli-paramagnetic character. The $\chi$ value of $\sim 3\times10^{-4}$ cm$^3$ mol-Ti$^{-1}$ is comparable to those of typical $d^1$ metallic compounds: $MgTi_2O_4$ [11], $VO_2$ [24], and $AVO_3$ ($A$ = Ca, Sr) [25], also indicating a Pauli paramagnetic metal. For comparison, $\chi$ of precursor $CuTi_2S_4$ is also shown. The $\chi$ value of $ZnTi_2S_4$ is nearly twice as large as its precursor $CuTi_2S_4$, which would be due to the electron doping. It has been previously reported that the $\chi$ value for Cu-deintercalated $Cu_{1-x}Ti_2S_4$ ($0 \leq x \leq 3/8$) decreases as the $x$ value increases [26]. This trend also supports the origin of $\chi$-increasing is the electron doping. In a low-temperature region, $\chi$ shows a Curie tail, which corresponds to 0.85% of $Ti^{3+}$ local moments ($S$ = 1/2). The Curie tail is tiny and sample-dependent, which strongly suggests it is extrinsic. Characteristically, the $\chi$ curve shows a broad peak approximately at $T_{max} \sim 110$ K (see also the inset), of which origin will be discussed later.

**IV. Discussion**

As described above, $ZnTi_2S_4$ is a thiospinel with a metallic ground state. First, we estimate the electron correlations in $ZnTi_2S_4$. The Wilson ratio $R_W = (\pi^2 k_B^2/3\mu_B^2)(\chi_0/\gamma) \sim 72.95(\chi_0/\gamma)$, which $\chi_0$ is $\chi$ at $T$ = 0, is often used to estimate the magnitude of electron correlations $U$ in itinerant electron systems [27]. For the free-electron Fermi gas, the value is expected to be $R_w$ = 1. Since both $\chi_0$ and $\gamma$ are proportional to the density of states and only $\chi_0$ is enhanced by electron correlations, $R_W$ increases from 1 to 2, respectively for the small and large limit of $U$. The $R_w$ value is roughly estimated to be approximately 1.6 with using the value of $\chi_0 \sim 3\times10^{-4}$ cm$^3$ mol-Ti$^{-1}$. On the other hand, the $R_w$ value of $CuTi_2S_4$ is approximately 1.2 from $\chi_0 \sim 2\times10^{-4}$ cm$^3$ mol-Ti$^{-1}$ and $\gamma$ = 11.95 mJ mol-Ti$^{-1}$K$^{-2}$. The tendency for the $\chi_0$-increase with the increase in the number of electrons has also been observed in $Cu_{1-x}Ti_2S_4$ ($0 \leq x \leq 3/8$) [26]. Thus, the electron correlation is enhanced by electron doping in the category of Pauli paramagnetism.

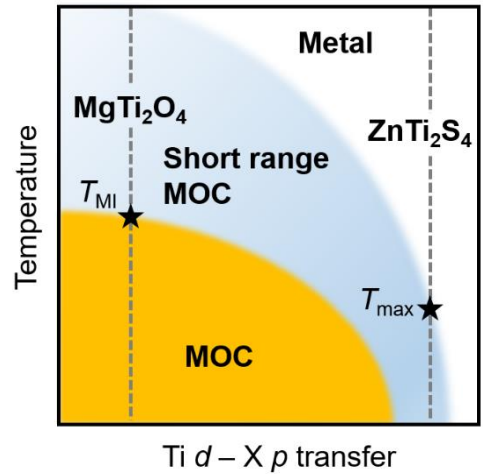

**Figure 4** Schematic phase diagram in $3d^1$ $Ti^{3+}$ spinel ($MgTi_2O_4$ and $ZnTi_2S_4$) system. The MOC and short-range MOC states are observed in the orange and blue regions, respectively.

The vast peak in $\chi$ at $T_{max}$ = 110 K is unusual for a paramagnetic metal. Namely, the $\chi$ value decreases below $T_{max}$. Considering the enhancement of $R_w$ value from $CuTi_2S_4$ to $ZnTi_2S_4$, this behavior would be attributed to electron correlation. Such a $\chi$-decreasing behavior is also observed in $MgTi_2O_4$ at higher temperatures than the MOC transition temperature $\sim$ 250 K [11]. Although the $\chi$ data in $MgTi_2O_4$ shows a monotonic decrease between 250 and 350 K, the presence/absence of broad peak has not been clarified since the magnetization measurements above 350 K have not been reported so far. Nevertheless, $MgTi_2O_4$ shows structural signatures of short-range Ti-Ti dimer formation up to at least 500 K demonstrated in x-ray and neutron experiments [28]. Thus, the observed vast peak of $\chi$ in $ZnTi_2S_4$ would be originated in a short-range fluctuation of MOC. The replacement of oxygen by sulfur would increase the overlap between the $3d$ orbital of Ti and the $p$ orbital (O: $2p$ and S: $3p$), which enhances itinerancy. We note that the $\chi$ data of $CuTi_2S_4$ seemingly approaches a constant value in the low-temperature region (shown by the chained line on the data), which also may be a signature of MOC fluctuation. Therefore, it is reasonable to think that $ZnTi_2S_4$ is an itinerant analog of $MgTi_2O_4$. The systematic evolution of electronic states from O to S is straightforward and summarized by a schematic phase diagram shown in Fig. 4. Such a change in the electronic state by anion substitution has also been observed in the triangular lattice compound $LiVS_{2-x}Se_x$ [4]. It would be possible to observe the metal-insulator transition associated with a MOC by controlling the electron kinetic energy by applying pressure or O substitution at the S site in ZnTi$_2$S$_4$: these are future issues.

In a broader context, we believe that there has been no report of the reaction using metallic zinc to exchange Cu$^+$ with Zn$^{2+}$ to prepare ZnTi$_2$S$_4$. This method is beneficial when the valence of the countercations can be controlled, as in the present case, which will allow the development of novel electronic properties. Furthermore, the cation exchange reaction allows for rational synthesis and targeting of the products, which is usually impossible in conventional high-temperature ceramic methods.

**V. Summary**

We successfully synthesized a novel thiospinel ZnTi$_2$S$_4$ via topochemical manipulation. Signature of the metallic ground state in ZnTi$_2$S$_4$ is observed in the Pauli-paramagnetic-like $\chi$, almost temperature-independent $\rho$, and a contribution of conduction electrons in *C*. Moreover, ZnTi$_2$S$_4$ shows a pseudogap-like behavior in $\chi$ below 110 K, which may be a precursor phenomenon of MOC. From these observations, we conclude that ZnTi$_2$S$_4$ is an itinerant analog of MgTi$_2$O$_4$ with a similar Ti$^{3+}$ pyrochlore lattice that exhibits MOC-driven metal-insulator transition. Our findings demonstrate that a newly-founded reductive topochemical reaction provides a promising platform for exploring a novel electronic state.

**Acknowledgments**

This work was supported by the Japan Society for the Promotion of Science (JSPS) KAKENHI Grants Nos. JP19K14646, JP18K03506, and JP21K03441. Part of this work was carried out by the joint research in the Institute for Solid State Physics, the University of Tokyo.